# FROM SU(2) GAUGE THEORY TO SPIN ½ QUANTUM MECHANICS

P. Zizzi
Department of Psychology, University of Pavia,
Piazza Botta, 6, 27100 Pavia, Italy
paola.zizzi@unipv.it

E. Pessa
Department of Psychology, University of Pavia,
Piazza Botta, 6, 27100 Pavia, Italy
eliano.pessa@unipv.it

**Abstract**
We consider a pure SU(2) gauge theory, and make an ansatz for the gauge field, which is gauge-invariant but manifestly non-Lorentz invariant. In a limit case of the ansatz, corresponding to a vacuum solution, the SU(2) gauge field reduces to a spin ½ observable times the generator of a global U(1).
We find that the field equations written in terms of the ansatz make explicit the presence of an anomalous current which vanishes in the vacuum. This allows to interpret the components of the U(1) field as Goldstone bosons associated with the spontaneous breaking of Lorentz symmetry.
Finally, we give an interpretation of the ansatz in the context of principal fiber bundles, which enlightens the geometrical aspects of the reduction of the gauge field theory to quantum mechanics.

# 1. Introduction

The SU(2) gauge theory was the first non-abelian generalization of the U(1) gauge theory of electromagnetism. It was introduced by Yang and Mills in 1954 [1], in order to extend the SU(2) global invariance of isotopic symmetry to a local SU(2) invariance. This requires the introduction of three vector fields, one for each generator of SU(2). These non-abelian gauge fields transform according to the adjoint representation of SU(2), and must be massless, since a mass term explicitly included in the lagrangian would spoil gauge invariance. Even pure SU(2) gauge theory is highly nonlinear, and the lagrangian contains self-interaction terms. The classical solutions of the field equations of pure SU(2) gauge theory have been extensively studied by a number of authors (for a review see, e.g., [2]). Because of the fact that the three vector fields should be associated with massless gauge bosons, pure SU(2) gauge theory was not considered as a theory of physical interest in itself. However, SU(2) gauge theory was exploited in theories where the Higgs mechanism gives masses to the gauge bosons, like in the Glashow-Salam-Weinberg model SU(2)x U(1) of electro-weak interactions [3], and the "standard model" SU(3)x SU(2)xU(1) unifying strong, weak and electromagnetic interactions (for a review see, e.g., [4]).

Nevertheless, we believe that a pure Yang-Mills theory (pure SU(2) gauge theory being the simplest case) can play a very important role in understanding how a gauge field theory and quantum mechanics are related to each other. The common opinion is that quantum field theory is just quantum mechanics plus special relativity. We don't disagree completely with that, but believe that there is something more. In fact, we will show that quantum mechanics can be obtained, through a suitable reduction mechanism, from a classical non-abelian gauge theory. This suggests that such a field theory, despite being classical, has a hidden quantum nature.

Of course, in this reduction process, the Lorentz invariance of the original gauge theory must be broken. Such a breaking is convenient because we want to get, as a result, quantum mechanics, which is not Lorentz invariant.

More specifically, the aim of this paper is to look for a mechanism which could reduce the SU(2) gauge theory to the quantum mechanics of spin ½.

The starting idea was to eliminate the infinite space-time degrees of freedom , which are those characterizing a field theory, and leaving only internal degrees of freedom.

This is realized by choosing a particular ansatz for the gauge field, which separates the dependence on space-time coordinates from that on internal degrees of freedom.

In this first step, the Lorentz symmetry is broken, but gauge invariance is preserved. In a second step, we look for a particular limit of the space-time dependent part of the ansatz, which leads to a vacuum solution of the field equations. In this limit, the full SU(2) gauge theory reduces to a quantum mechanical theory.

The geometrical description of such a reduction is given in terms of a local section of the principal fiber bundle, which becomes constant due to a contraction mapping related to the ansatz.

The paper is organized as follows.

In Sect. 2, we make an ansatz for the SU(2) gauge field, in terms of the exponential of a U(1) gauge field times a Pauli matrix. This breaks Lorentz invariance, but not gauge invariance.

In the limit case where the U(1) gauge field tends to zero, the ansatz describes a new vacuum solution.

In Sect. 3, we consider the field equations written in terms of the ansatz, and find an anomalous current, due to the self-interaction of the SU(2) gauge field, put in evidence by the ansatz itself.

In Sect. 4, we show that, in the vacuum, the U(1) gauge field in the ansatz describes free massless particles which are interpreted as the Goldstone bosons associated with the breaking of Lorentz invariance.

In Sect. 5, we consider the principal fiber bundle for the SU(2) gauge theory, and make a particular choice for the U(1) gauge field in the ansatz, in such a way that the centre of an open ball in the base space topology is an attractive fixed point. The U(1) gauge field is then a contraction mapping in the basin of the attractor.

In Sect. 6, we show that the contraction mapping acts on the open covering of the base space in such a way that the local sections become constant, and the principal fiber bundle becomes trivial. The principal connection vanishes, and the SU(2) gauge field reduces to the generator of a global U(1) times a Pauli matrix.

Sect. 7 is devoted to the conclusions.

## 2. The ansatz

We consider the SU(2) gauge field $A_\mu^a(x)$ ($\mu = 0,1,2,3$; $a = 1,2,3$) and make the following ansatz:

$$A_\mu^a(x) = e^{-i\lambda_\mu(x)} \sigma^a \qquad (2.1)$$

where $\lambda_\mu(x)$ is a U(1) gauge field and the $\sigma^a$ are the Pauli matrices, which satisfy the commutation relations:

$$[\sigma^a, \sigma^b] = 2i\varepsilon_{abc}\sigma^c \qquad (2.2)$$

The ansatz (2.1) explicitly breaks Lorentz invariance.

In the following we will consider, in particular, the limit case:

$$\lambda_\mu(x) \to 0 \qquad (2.3)$$

In this limit one gets $A_\mu^a(x) \to \sigma^a$. In a sense, the SU(2) gauge theory reduces to the quantum mechanics of spin ½.

Let us consider the SU(2) gauge transformations performed on the original gauge field:

$$A_\mu \equiv A_\mu^a \sigma^a / 2 \qquad (2.4)$$

that is

$$A_\mu \xrightarrow{U} A_\mu' = U A_\mu U^{-1} - \frac{i}{g} U \partial_\mu U^{-1} \qquad (2.5)$$

where $g$ is the gauge coupling constant, $U$ is given by:

$$U = \exp(i\rho^a(x)\sigma^a / 2) \qquad (2.6)$$

and $\rho^a(x)$ are three arbitrary real functions.

By the use of (2.4) the ansatz (2.1) can be rewritten as:

$$A_\mu = e^{-i\lambda_\mu} \qquad (2.7)$$

The ansatz (2.7) transforms under (2.5) as:

$$e^{-i\lambda_\mu} \xrightarrow{U} e^{-i\lambda_\mu'} = e^{-i\lambda_\mu} - \frac{i}{g} U \partial_\mu U^{-1} \qquad (2.8)$$

In the limit case (2.3) the transformations (2.8) become:

$$e^{-i\lambda_\mu} \xrightarrow{U} e^{-i\lambda_\mu'} = 1 - \frac{i}{g} U \partial_\mu U^{-1} \qquad (2.9)$$

Eq. (2.9) can be transformed into a pure gauge by a suitable choice of the arbitrary functions $\rho^a(x)$. This means that in the limit case the ansatz (2.1) describes a vacuum solution.

In the original SU(2) theory invariant under Lorentz transformation, the vacuum state was $|0\rangle$, corresponding to $A_\mu = 0$. In presence of the ansatz, which breaks Lorentz invariance, there is, in the limit case, a new vacuum state $|\vartheta\rangle$, corresponding to $A_\mu = 1$.

Then, the gauge field $A_\mu$ has a non-vanishing v.e.v. in the new vacuum:

$$\langle \vartheta | A_\mu | \vartheta \rangle \neq 0 \qquad (2.10)$$

Let us take the temporal gauge $A_0 = 0$. Then, Eq. (2.10) becomes:

$$\langle \vartheta | A_i | \vartheta \rangle \neq 0 \qquad (i = 1, 2, 3) \tag{2.11}$$

This indicates that there is a spontaneous symmetry breaking of the little group O(3), to which are associated three Goldstone bosons $\varphi_i$, each one corresponding to a particular O(3) generator.

In the case that the Lorentzian Goldstone bosons were not self-interacting, they would obey the field equations $\Delta \varphi_i = 0$, where $\Delta$ is the Laplacian. However, as we will see in the next section, this is not the case. In fact, to satisfy the original SU(2) field equations, the Lorentzian Goldstone bosons must be the source of a current:

$$\Delta \varphi_i = j_i \qquad (i = 1, 2, 3) \tag{2.12}$$

## 3. The anomalous current

Let us consider the SU(2) gauge field in the notation (2.4).

In this notation, the expressions for the tensor field, for the covariant derivative, for the lagrangian density, and for the field equations are, respectively, given by:

$$F_{\mu\nu} = \partial_\mu A_\nu - \partial_\nu A_\mu + ig [A_\mu, A_\nu] \tag{3.1}$$

$$D_\mu = \partial_\mu + ig A_\mu \tag{3.2}$$

$$L = -\frac{1}{4} F_{\mu\nu} F^{\mu\nu} \tag{3.3}$$

$$D_\mu F_{\mu\nu} = 0 \tag{3.4}$$

In terms of the ansatz (2.7), equations (3.1), (3.2) and (3.3) take, respectively, the form:

$$F_{\mu\nu} = i(e^{-i\lambda_\mu} \partial_\nu \lambda_\mu - e^{-i\lambda_\nu} \partial_\mu \lambda_\nu) \tag{3.5}$$

$$D_\mu = \partial_\mu + ig e^{-i\lambda_\mu} \tag{3.6}$$

$$L = \frac{1}{4} \left[ e^{-2i\lambda_\mu} (\partial_\nu \lambda_\mu)^2 + e^{-2i\lambda_\nu} (\partial_\mu \lambda_\nu)^2 - 2 e^{-i(\lambda_\mu + \lambda_\nu)} \partial_\nu \lambda_\mu \partial_\mu \lambda_\nu \right] \tag{3.7}$$

Notice that in (3.7) there is a "mass term" $-\frac{1}{2} A_\mu A_\nu \partial_\nu \lambda_\mu \partial_\mu \lambda_\nu$ for the SU(2) gauge field $A_\mu$ in the ansatz (2.7).

The conserved Noether current $j^N{}_\nu$ is, in terms of the ansatz:

$$j^N{}_\nu = \frac{\partial L}{\partial (e^{-i\lambda_\nu})} = e^{-i\lambda_\nu} \left[ (\partial_\mu \lambda_\nu)^2 - \partial_\nu \lambda_\mu \partial_\mu \lambda_\nu \right] \tag{3.8}$$

Finally, the field equations written in terms of the ansatz become:

$$e^{-i\lambda_\mu} \partial_\mu \lambda_\mu \partial_\nu \lambda_\mu - e^{-i\lambda_\nu} (\partial_\mu \lambda_\nu)^2 + i e^{-i\lambda_\mu} \partial_\nu (\partial_\mu \lambda_\mu) - i e^{-i\lambda_\nu} \partial_\mu^2 \lambda_\nu$$
$$- g e^{-i\lambda_\mu} (e^{-i\lambda_\mu} \partial_\nu \lambda_\mu - e^{-i\lambda_\nu} \partial_\mu \lambda_\nu) = 0 \tag{3.9}$$

In the Lorentz gauge $\partial_\mu \lambda_\mu = 0$, the first and third terms in (3.9) vanish, and the field equation can be rewritten as:

$$\Box e^{-i\lambda_\mu} = j_\mu \tag{3.10}$$

where $\Box$ is the d' Alembertian, and $j_\mu$ is the current:

$$j_\mu = g e^{-i\lambda_\mu} \left( e^{-i\lambda_\mu} \partial_\nu \lambda_\mu - e^{-i\lambda_\nu} \partial_\mu \lambda_\nu \right) \tag{3.11}$$

which is not conserved:

$$\partial_\mu j^\mu = g e^{-i(\lambda_\mu + \lambda_\nu)} \left[ i(\partial_\mu \lambda_\nu)^2 - \partial_\mu^2 \lambda_\nu \right] \tag{3.12}$$

Notice that the ansatz (2.7) puts in evidence the fact that the anomaly (3.12) is due to the self-interaction of the SU(2) gauge field.

## 4. The Lorentzian Goldstone bosons

The last term in (3.9) describes a self-interaction of the U(1) gauge field $\lambda_\mu$.

In the vacuum:
$$\lambda_\mu = 0 \ , \qquad A_\mu = 1 \tag{4.1}$$
the self-interaction must be absent, so that the current (3.11) vanishes:
$$j_\mu = 0 \tag{4.2}$$
From (3.11) and (4.2) it follows that in the vacuum it holds:
$$\partial_\nu \lambda_\mu = \partial_\mu \lambda_\nu = 0 \tag{4.3}$$
Under these conditions, also the first and second terms in (3.9) vanish. Moreover, we recall that, in the Lorentz gauge, also the third term vanishes.

Then, in order to satisfy the SU(2) gauge field equations, the fourth term must be equal to zero:
$$\lambda_\nu = 0 \tag{4.4}$$
Moreover, notice that there is no mass term of the kind $\lambda_\mu \lambda_\mu$ for the U(1) field in the lagrangian density (3.7), which reflects in the absence of a linear term in $\lambda_\mu$ in the field equations (3.9).

Then, in the vacuum the $\lambda_\mu$ fields describe massless free particles like the Goldstone bosons.

Therefore, once the gauge has been fixed, we interpret the three components of the U(1) field $\lambda_\mu$ as the three Lorentzian Goldstone bosons $\varphi_i$.

## 5. The contraction mapping

The pure SU(2) gauge theory under consideration can be described in terms of a principal fiber bundle $(P, \pi, B, G)$ [5] where $P$ is the total space, $B$ is the base space (in our case $R^4$), $G$ (in our case SU(2)) is the structure group, which is homeomorphic to the fiber space $F$, and $\pi$ is the canonical projection:
$$\pi : P \to R^4 \tag{5.1}$$
The base space $R^4$ is equipped with the Euclidean metric $d$:
$$d(x', x) = |x' - x| \tag{5.2}$$
where $x$ and $x'$ are two points of $R^4$ and must be intended as $x \equiv \{x_\mu\}$, $x' \equiv \{x_\mu'\}$ $(\mu = 1, 2, 3, 4)$.

The complete metric space $(R^4, d)$ has an induced topology which is that of the open balls with rational radii $r_n = \dfrac{1}{n}$, with $n$ a positive integer.

The open ball of rational radius $r_n$, centred at $x^*$ is:
$$B_{r_n}(x^*) = \left\{ x \in R^4 \,\middle|\, d(x^*, x) < r_n \right\} \tag{5.3}$$
The set of open balls $B_{r_n}(x^*)$ is an open covering of $R^4$ and forms a local basis for the topology.

Now, let us consider again the ansatz (2.7), and make the following natural choice for $\lambda_\mu(x)$:
$$\lambda(x) = x^* e^{i \frac{|x^* - x|}{n}} \tag{5.4}$$
where $\lambda$ in (5.4) must be intended as $\lambda \equiv \{\lambda_\mu\}$ $(\mu = 1, 2, 3, 4)$.

The point $x^*$ is a fixed point for $\lambda(x)$ as it holds:
$$\lambda(x^*) = x^* \tag{5.5}$$
It is easy to check that $\lambda(x)$ continuously approaches $x^*$ for large values of $n$ (i.e., for smaller radius of the ball):
$$\lim_{n \to \infty} \lambda(x) = x^* . \tag{5.6}$$

The fixed point $x^*$ is an *attractive* fixed point for $\lambda(x)$, as it holds:
$$|\lambda'(x^*)| < 1 \tag{5.7}$$
The point $x^*$ is then a particular kind of attractor for the dynamical system described by this theory. Furthermore, it holds:
$$|\lambda'(x)| < 1 \tag{5.8}$$
for all $x \in B_{r_n}(x^*)$, which is equivalent to say that $\lambda(x)$ is a contraction mapping in the attraction basin of $x^*$, that is, it satisfies the Lipschitz condition [6]:
$$d(\lambda(x), \lambda(x')) \leq q\, d(x, x') \tag{5.9}$$
with $q \in (0,1)$ for every $x, x' \in B_{r_n}(x^*)$.

## 6. Global trivialization induced by the contraction mapping

Let us denote the fiber over the attractive fixed point $x^*$ by:
$$F_{x^*} \equiv \pi^{-1}(x^*) \tag{6.1}$$
In this Section, we will show that, due to the contraction mapping, all fibers $F_x \equiv \pi^{-1}(x)$ coincide with (6.1) and with the abstract fiber $F \cong SU(2)$ for every $x \in R^4$, giving rise to the trivial bundle $\pi: R^4 \times SU(2) \to R^4$.

Then, the SU(2) principal connection vanishes and the gauge field reduces to the generator of a global U(1) group times a Pauli matrix.

Let us consider the principal fiber bundle (5.1) with a local trivialization $\{\varphi_i, U_i\}$, where $\varphi_i$ is the diffeomorphism:
$$\varphi_i : \pi^{-1}(U_i) \to U_i \times SU(2) \tag{6.2}$$
and the open neighbourhood $U_i$ is the open ball (5.3), which can be expressed as:
$$U_i(x) = \left\{ x \,\middle|\, |x^* - x| < \frac{1}{n} \right\} \tag{6.3}$$
In (6.2) the map $\varphi_i$ is defined as:
$$\varphi_i(\pi^{-1}(x)) = (x, g) \tag{6.4}$$
for every $x \in U_i$ and $g \in SU(2)$.

The canonical local section associated with the local trivialization $\{\varphi_i, U_i\}$ is defined as:
$$s_i(x) \equiv \varphi_i^{-1}(x, e) \tag{6.5}$$
where $e$ is the identity element of SU(2), and it holds:
$$\pi(s_i(x)) = x \qquad \text{for every } x \in U_i \tag{6.6}$$
Now, let us express the open neighbourhood $U_i$ in (6.3) in terms of the contraction mapping (5.4):
$$U_i(\lambda(x)) = \left\{ \lambda(x) \,\middle|\, -i \ln\left(\frac{\lambda(x)}{x^*}\right) < \frac{1}{n^2} \right\} \tag{6.7}$$
For $n \to \infty$ we have $\lambda(x) \to x^*$, and the open neighbourhood $U_i$ becomes the singlet:
$$U_i = \{\|x^*\|\} \tag{6.8}$$
Then, because of the contraction mapping, all fibers over $x \in U_i$ become:
$$\pi^{-1}(x) \equiv \pi^{-1}(x^*) \tag{6.9}$$
The local trivialization (6.4) becomes:
$$\varphi_i(\pi^{-1}(x^*)) = (x^*, g) \tag{6.10}$$

and the local section (6.5) becomes a constant section:
$$s_i(x) \equiv \varphi_i^{-1}(x^*, e) = s_i(x^*) \equiv \bar{s}_i \tag{6.11}$$

The most natural choice of an atlas, in this case, is to take all the local trivialization charts of the same kind of $\{\varphi_i, U_i\}$. Let us consider for example a second chart $\{\varphi_j, U_j\}$, where $U_j$ is the open ball centred at $x^{*'}$, which can be rewritten in terms of a contraction mapping:

$$\lambda(x)' = x^{*'} e^{i\frac{|x^{*'} - x|}{n}} \tag{6.12}$$

and $\varphi_j$ is a local trivialization of the same kind of (6.2).

The canonical local section $s_j$ associated with the local trivialization $\{\varphi_j, U_j\}$ is defined in the same way as (6.5). For $n \to \infty$ we have $\lambda(x) \to x^{*'}$, and the open neighbourhood $U_j$ becomes the singlet:

$$U_j = \{\|x^{*'}\|\} \tag{6.13}$$

Then, because of the contraction mapping, all fibers over $x \in U_j$ become:

$$\pi^{-1}(x) \equiv \pi^{-1}(x^{*'}) \tag{6.14}$$

The local trivialization becomes:
$$\varphi_j(\pi^{-1}(x^{*'})) = (x^{*'}, g) \tag{6.15}$$

and the local section $s_j$ becomes a constant section:
$$s_j(x) \equiv \varphi_j^{-1}(x^{*'}, e) = s_j(x^{*'}) \equiv \bar{s}_j \tag{6.16}$$

So that we have:
$$s_i(x) = \begin{cases} s_i(x^*) & \text{for } x = x^* \\ 0 & \text{for } x \neq x^* \end{cases} \qquad \text{for} \qquad U_i = \{\|x^*\|\} \tag{6.17}$$

$$s_j(x) = \begin{cases} s_j(x^{*'}) & \text{for } x = x^{*'} \\ 0 & \text{for } x \neq x^{*'} \end{cases} \qquad \text{for} \qquad U_j = \{\|x^{*'}\|\} \tag{6.18}$$

We recall that the general relation between two local sections $s_i$ and $s_j$, canonically associated respectively with the local trivializations $\varphi_i$ and $\varphi_j$, is:

$$s_j(x) = s_i(x) t_{ij}(x) \qquad \text{for every } x \in U_i \cap U_j \tag{6.19}$$

where $t_{ij}(x)$ are the transition functions, which are defined by:
$$t_{ij}(x) = \varphi_i(x) \circ \varphi_j^{-1}(x) \tag{6.20}$$

Let us consider the two possible cases:
i) $x^* \in U_i \cap U_j$
ii) $x^{*'} \in U_i \cap U_j$

Let us consider first case i). From (6.19) we have:
$$s_j(x^*) = s_i(x^*) t_{ij}(x^*) \tag{6.21}$$
with:
$$s_j(x^*) = 0 \tag{6.22}$$
because of (6.18).

In the same way, in case ii) we get:
$$s_j(x^{*'}) = s_i(x^{*'}) t_{ij}(x^{*'}) \tag{6.23}$$
with:

$$s_i(x^{*\prime}) = 0 \tag{6.24}$$

because of (6.17).
From the above we get:

$$s_j(x^{*\prime}) = s_i(x^*)t_{ij}(x^*) \tag{6.25}$$

which is consistent only for $x^* = x^{*\prime}$ and $i = j$, leading to the identity:

$$s_i(x^*) = s_i(x^*) \tag{6.26}$$

as it is $t_{ii}(x^*) = 1$.

This means that if we take an atlas whose local trivialization charts are associated with the same contraction mapping, the contraction point is unique.

We discard the situation where some open neighbourhoods were contracted to their centre and other were not, as the space-time base manifold would become disconnected, and in this case the fixed point attractor $x^*$ would be in fact a singularity.

The principal connection is defined as:

$$A_i \equiv s_i^* \omega \tag{6.27}$$

where $s^*$ is the pullback of the local section (6.5) and $\omega$ is a one-form defined on P:

$$\omega : T_p P \to \mathcal{G} \tag{6.28}$$

In (6.28) $T_p P$ indicates the tangent space of the total space $P$ at point $p \in P$, and $\mathcal{G}$ stands for the SU(2) algebra.

In our case, due to the contraction mapping, we have to consider the pullback of the constant section (6.11) and replace Eq. (6.27) with:

$$A_i \equiv \bar{s}_i^* \omega \tag{6.29}$$

We recall that the relation between the connection $A_i$ in (6.11) and the gauge field $A_\mu$ is:

$$A_i \equiv -ig A_\mu^a \sigma^a dx_\mu \tag{6.30}$$

where $g$ is the SU(2) coupling constant, $\sigma^a$ are the Pauli matrices, and $x_\mu$ are local coordinates in $U_i$.

By inserting the ansatz (2.1) in (6.30) we get:

$$A_i \equiv -ig e^{-i\lambda_\mu(x)} dx_\mu \tag{6.31}$$

for every $x \in U_i$.

In correspondence to the contraction point $x^*$, the connection coefficients in (6.31) become $-ig e^{-ix^*}$. However, the connection $A_i$ vanishes because $dx_\mu = 0$.

The SU(2) gauge field $A_\mu^a$ reduces to an operator $A^a$ which, up to a multiplicative constant, is the product of the generator of a global U(1) group times a Pauli matrix:

$$A^a = -ig e^{-ix^*} \sigma^a \tag{6.32}$$

This means that the pure SU(2) gauge field theory is reduced to a quantum mechanical theory of spin ½ with a constant U(1) "charge", in absence of any interaction.

## 7. Conclusions

In this paper, we described the reduction of the pure SU(2) gauge theory down to the quantum mechanics of spin ½ in terms of an ansatz for the gauge field in the vicinity of an attractive fixed point. This suggests that qubits might be generated in an attractor basin of the SU(2) gauge theory. Then, quantum information theory might have roots in a classical non-abelian gauge theory.

This result seems to indicate that quantum mechanics can be the outcome of a classical gauge theory only when the gauge group is non-abelian. Roughly speaking, the presence of a non-abelian group in a classical field theory signals the presence of a hidden quantum nature.

It might be interesting to apply the above procedure to a gauge theory of gravity. However, most classical gauge theories of gravity are abelian, thus they don't have a hidden quantum nature. Among these theories, a typical example is the Einstein-Cartan-Sciama-Kibble (ECSK) theory [7] [8] based on the Poincaré gauge group. Most probably, the absence of a hidden quantum nature in ECSK, due to the abelian character of the Poincaré group (and in other theories of this kind) could be at the root of the difficulties encountered when looking for a quantum theory of gravity.

In this regard, one should consider, instead, classical non-abelian gauge theories of gravity, like, for instance, the general diffeomorphism invariant SU(2) gauge theory of gravity [9].

The fact that the reduction described above is possible only when all the local trivialization charts are associated with contraction mappings, unless a disconnected space manifold is taken into account, means that we cannot have a general theory which describes at once a classical gauge theory and quantum mechanics in a smooth space-time manifold. In fact, black holes, for which a quantum mechanical theory has been formulated [10], although they arise as classical solutions of general relativity, show that the coexistence of field theory and quantum mechanics is possible only at the expenses of smoothness, as the singularity in their interiors makes space-time disconnected.

As we have seen, once one requires that all the local trivialization charts are associated with contraction mappings, the attractor is unique. If instead we would require multiple attractors, we should choose those contraction mappings such that the size scale of the attractor basins is the Planck scale. This would give rise to a multiple reduction of the original field theory in differents regions of space-time. In the case of SU(2), the scenario would be that in which each Planckian cell encodes one qubit. This is a possible approach to loop quantum gravity [11] in terms of quantum-computational space-time [12].